\documentclass{pasj00}

\begin{document}
\SetRunningHead{Fleishman et al.}{Relativistic Positrons}

\title{Microwave Signature of Relativistic Positrons in Solar Flares}

\author{Gregory \textsc{Fleishman}}
 \affil{Center For Solar-Terrestrial Research,
New Jersey Institute of Technology, Newark, NJ 07102, USA}
\email{gfleishm@njit.edu}
\author{Alexander \textsc{Altyntsev}}
\and
\author{Nataliia \textsc{Meshalkina}}
 \affil{Institute of Solar-Terrestrial
Physics SB RAS, Lermontov St. 126A, Irkutsk 664033, Russia}
\email{altyntsev@iszf.irk.ru, nata@iszf.irk.ru}



\maketitle

\begin{abstract}
Relativistic antiparticles can be created in high-energy nuclear
interactions; thus, detection of antiparticles in an astrophysical
source can tell us something remarkable about the underlying
high-energy processes and nuclear interactions. However, once
created, the antiparticles remain a minor fraction of their
conjugant normal particles, so the detection of the antiparticles
represents a big science challenge. To address this challenge we
employ imaging and polarimetry of microwave radiation produced as
the positrons gyrate in the ambient magnetic field. The key
property of the radiation used in this method is that the
oppositely charged particles, electrons and positrons, produce
radiation with opposite helicity, easily distinguishable by
currently operating radio facilities. Analysis of available
spatially resolved microwave data augmented by independent
magnetic field measurements allows us to remotely detect the
relativistic positron component in several solar flares.

\end{abstract}

\section{Introduction}

One of the puzzling mysteries of the world around us is a strong
asymmetry between the matter and antimatter (\cite{Defouw1970},
\cite{Stecker}). Under normal conditions, the antimatter is almost
absent \citep{Stecker}, although relativistic antiparticles can be
created in high-energy nuclear interactions \citep{Eichler, Murphy}. Thus,
detection of antiparticles in an astrophysical source can tell us
something remarkable about the underlying high-energy processes
and nuclear interactions \citep{Ramaty}. However, since the
antiparticles constitute at most a minor fraction of their
conjugant normal particles \citep{Share}, detection of the
antiparticles represents a big science challenge. Currently, there
are no direct means of remote diagnostics of \emph{relativistic} antiparticles.
Here we report on the first detection of a relativistic positron signature in
the polarized microwave solar spectra employing the fact that the oppositely
charged particles, electrons and positrons, produce radiation with
opposite helicity \citep{Melrose}. Thus, the key observable of the
radiation used in this method is its circular polarization, easily
measurable by currently operating radio facilities.

In the classical, Nobel-Prize-winning experiments by Carl Anderson
\citep{Anderson}, positrons were identified in the laboratory by
tracking their trajectories in the magnetic field. Having similar
shapes to the electron trajectories, the positron trajectories
spiraled in the opposite direction. In fact, they appeared as a
reflection of an electron trajectory, revealing a particle of
positive charge with properties otherwise identical to the
electron's. In our study we also employ this mirror symmetry of
the trajectories. The electron and positron moving along such
oppositely twisted paths emit electromagnetic radiation with the
same intensity, but opposite circular polarization helicity
\citep{Melrose}. The circular polarization is an easily measurable
quantity in the microwave range and so offers an elegant means of
remote diagnostics of positrons in astrophysical sources. Here we
apply this simple polarimetry method for analysis of microwave
emission from the Sun and report on the first detection of
relativistic positrons produced in solar flares presenting ample
evidence for the relativistic positron discovery in one of them.

We have to note that production of the positrons in solar flares
is well established via observation of the positron-electron
annihilation line at 0.511 MeV (\cite{Chupp_etal_1973},
\cite{Share}, \cite{Schrijver}). This
line, however, is formed by annihilation of already thermalized
slow, nonrelativistic positrons, after they have interacted with a
substantial column depth of the ambient plasma \citep{Murphy}.
Thus, conclusions about the site where the positrons are
originally produced, their transport and escape, made based on the
annihilation line, are uncertain.

In contrast, relativistic positrons are highly efficient emitters
of the bremsstrahlung due to collisions with ambient particles and
gyro emission due to spiraling in the magnetic field
(\cite{Adriana}, \cite{Trottet}, \cite{Krucker}) and so start to
radiate immediately once created. The challenge is to distinguish
the radiation produced by positrons from that by electrons. In
particular, a broad gamma-ray feature with a maximum near 70~MeV
appears in the high-energy gamma-ray spectrum of some flares. By
now this feature has been reported in about 20 flares (e.g.,
\cite{Forrest85}, \yearcite{Forrest86}; \cite{Chupp};
\cite{Akimov}, \yearcite{Akimov94}, \yearcite{Akimov96};
\cite{Grechnev08}; \cite{Kuznetsov}; \cite{Kurt}). This emission
is caused by the decay of neutral pions ($\pi^0$), which are
produced in interactions of high-energy protons ($>300$~MeV) with
dense layers of the solar atmosphere (\cite{Murphy84};
\cite{Ramaty87}; \cite{Murphy87}; \yearcite{Murphy2007}). Charged
pions, $\pi^\pm$, preferentially $\pi^+$, are also produced in the
very same interactions, and they then decay into electrons and
(preferentially) positrons, which, in turn, generate high-energy
bremsstrahlung with a broad energy spectrum extending up to the
energies of parent electrons and positrons. But the gamma-ray
bremsstrahlung produced by these secondary positrons is
observationally indistinguishable from the bremsstrahlung produced
by electrons with the same spectrum. In addition, in some cases a
purely leptonic fit to the observed gamma-ray spectrum may have
the goodness of fit comparable to that of the hybrid (leptonic
plus hadronic) fit \citep{Ackermann12}, which raises an
ambiguity of the data interpretation. Remarkably, we can remove
this ambiguity studying the gyro emission in the microwave range,
where polarization of emission is available in addition to the
radiation intensity.

\section{Employed Methodology}

A single electron gyrating in a magnetic field produces
electromagnetic radiation with the predominant polarization in the
sense of extraordinary mode, while a single positron produces
ordinary mode radiation \citep{Melrose}. They are, respectively,
the right-circular polarization (RCP) and left-circular
polarization (LCP) for the line of sight magnetic field component
directed to the observer, and vice versa for the oppositely
directed magnetic field. Thus, the key parameter needed to
distinguish gyrosynchrotron radiation from electrons and positrons
is the sense of circular polarization relative to the direction of
the line of sight component of the magnetic field at the source.

In the astrophysical context, including solar flares, the problem
is complicated by the fact that radiation is produced by particle
ensembles with broad energy distributions \citep{Lin03}. Fast
electrons and protons are produced in flares by acceleration of
low-energy thermal particles (\cite{Shih}, \cite{Lin11}) and can
be very numerous \citep{Fleishman}. In contrast, fast positrons
are created as highly energetic \textit{relativistic} particles as
a byproduct of nuclear interactions of the accelerated ions
(\cite{Kozlovsky}, \cite{Murphy}, \cite{Ramaty}, \cite{Charles}).
Specifically, the positrons can arise from either $\beta^+$ decay
of the flare-produced radioactive nuclei (\cite{Ramaty},
\cite{Adriana}), $N \to N' + e^{+} + \nu_e$, where N is the
original radioactive nucleus, e.g. $^{11}\mathrm{C}$ or
$^{15}\mathrm{O}$, N' is the final nucleus, $e^{+}$ is the
positron and $\nu_{e}$ is the electron neutrino, or decay of
$\pi^{+}$ particles (\cite{Charles}, \cite{MacKinnon}), $\pi^{+}
\to \mu^{+} + \nu_{\mu}$, $\mu^{+} \to e^{+} + \nu_{e} +
\overline{\nu_{\mu}}$, where $\mu^{+}$ is anti mu--meson. The
direct decay, $\pi^{+} \to e^{+} + \nu_{e}$, is also possible but
less probable. Finally, some excitation of nuclear level decay
with creation of the electron-positron pair (\cite{Kozlovsky}).
This means that a positron component must be recognized on top of
typically more numerous fast electrons.
This is a non--trivial task in a general case although it can be
feasible under some favorable combination of parameters.

\begin{figure}
  \begin{center}
    \FigureFile(85mm,85mm){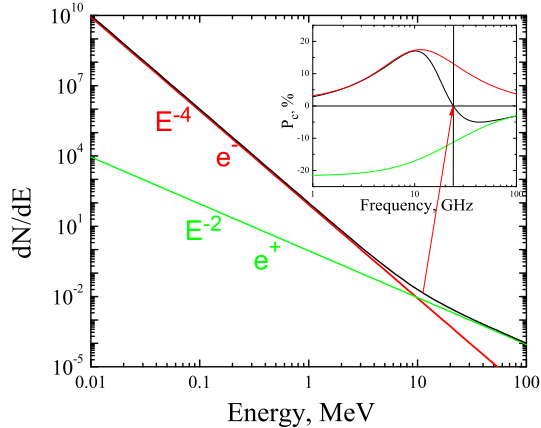}
  \end{center}
  \caption{ A cartoon illustrating the approach employed to distinguish the
positron contribution. Schematic energy spectra of fast electrons
(red) and positrons (green) favorable for the high-energy positron
detection. Black line shows the composite spectrum dominated by
electrons at low energies, but by positrons at high energies. The
total number of positrons is more than 5 orders of magnitude
smaller than the number of electrons. Behavior of the circular
polarization of the emission produced by these two components and
by their composition is shown in the inset. A distinctive
signature of the positron presence is the polarization reversal at
a high frequency shown by red arrow.}\label{fig:Fig1}
 \end{figure}

Figure \ref{fig:Fig1} envisions such a favorable situation and
illustrates the proposed method of fast positron detection. Let us
consider a typical situation when the total number of fast
positrons is only a minor fraction of the total number of fast
electrons, but the electron energy spectrum (red line in
Fig.~\ref{fig:Fig1}) is steeper than the positron energy spectrum
(green line; note that this power-law spectrum does not directly refer to any particular model of the positron origin at the source of the gyro emission). The composite spectrum (black line) is dominated by
the electrons at low energies, below 10 MeV in our example, while
the positrons dominate at higher energies (\cite{Charles},
\cite{MacKinnon}). Accordingly, the polarization sense of the
radiation produced by such a composite particle spectrum will
change at some frequency, as shown by the black curve in the inset
in Fig. \ref{fig:Fig1}. This frequency value depends on the
magnetic field strength at the source and can vary significantly
from one event to another. In this example, for the magnetic field
about 100~G, the polarization changes the sign around 25~GHz. In
contrast, no polarization reversal takes place if no positron
component is present (the red curve in the inset).

Thus, the search for a positron contribution to the microwave
emission from solar flares must rely on imaging observations at a
sufficiently high microwave frequency. This imaging is needed to
locate the source of emission and reliably specify the
corresponding emission mode (ordinary or extraordinary).
Fortunately, there exists a unique solar-dedicated radio
instrument that has already accumulated a significant database of
solar microwave bursts at the required spectral range, the
Nobeyama Radioheliograph (NoRH) \citep{Nakajima94}, that is
perfectly suited for our study. NoRH produces images of intensity
($I=R+L$) and polarization ($V=R-L$) at 17 GHz while of the
intensity only at 34 GHz. In addition, Nobeyama Polarimeters
(NoRP) \citep{Nakajima85} observe total power data (both I and V)
at a number of single frequencies including 17 and 35 GHz. This
set of observational tools suggests the following strategy of
identifying properties of solar bursts with unambiguous positron
contribution: (i) single, spatially coinciding, sources at both 17
and 34 GHz; (ii) the 34 GHz emission must come from an area where
the 17 GHz V displays a unipolar distribution (i.e., the
polarization of 17 GHz emission has a definite sense throughout
the region of 34 GHz emission); and (iii) the total power V must
have opposite signs at 17 and 34 GHz. This combination of observed
properties guarantees that we observe extraordinary mode radiation
(produced by electrons) at 17 GHz, but ordinary mode radiation
(produced by positrons) at 34 GHz. Finally, the identification of
the extraordinary mode at 17 GHz must be confirmed by comparison
of its polarization helicity with the direction of the
line--of--sight magnetic field independently available from
high--resolution optical spectro-polarimetry \citep{Scherrer}.
This is needed to firmly exclude two other competing mechanisms
capable of forming ordinary mode radiation from electrons --
polarization reversal due to either mode coupling or beam--like
pitch--angle anisotropy -- as either of these mechanisms, if
operational at 34 GHz, would also change the sense of polarization
at lower frequencies including 17 GHz.

\section{Database search and event analysis}

We searched the entire Nobeyama database for the described
positron signature. As in Anderson's classical experiment
\citep{Anderson}, where about 1\% of all tracks were produced by
positrons (15 tracks of 1300 recorded tracks), in about 1\% (6 of
600+) of all microwave bursts recorded by the Nobeyama
instruments, NoRH and NoRP, we have confidently identified a
contribution produced by the relativistic positrons. Let us
consider one burst that occurred on 13 March 2000 in more detail.

\begin{figure}
  \begin{center}
    \FigureFile(85mm,85mm){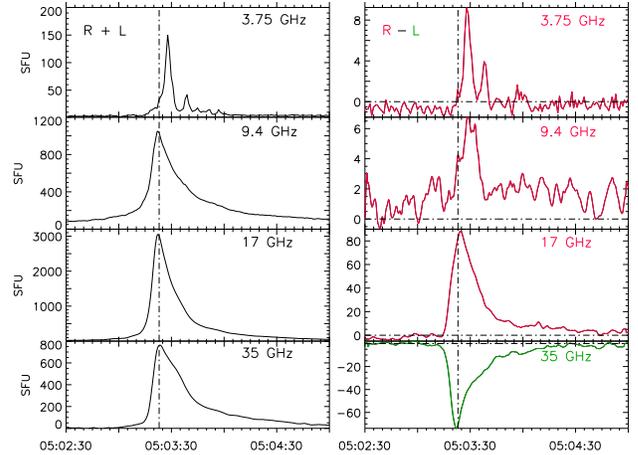}
  \end{center}
  \caption{Microwave light curves for 13 Mar 2000 event. Total intensity ($I=R+L$),
left column, and polarization ($V=R-L$), right column, from NoRP
at four different frequencies as indicated in the panels. RCP
emission, presumably produced by electrons is shown in red; LCP
emission, presumably produced by positrons is shown in green. 1
SFU = $10^{-22}$ Watt m$^{-2}$ Hz$^{-1}$ is the Solar Flux Unit; time hh:mm:ss
is the Universal Time, UT.}\label{fig:Fig2}
 \end{figure}

\begin{figure}
  \begin{center}
    \FigureFile(85mm,85mm){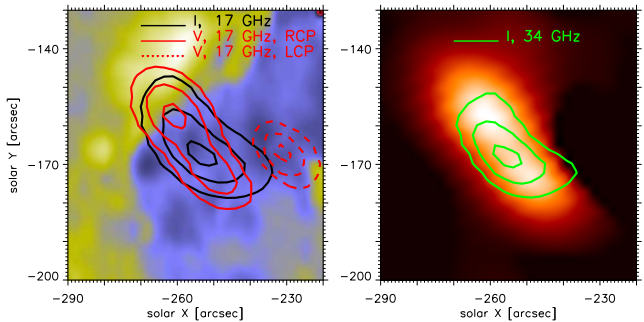}
  \end{center}
  \caption{Imaging data. Left panel shows color image of the
line-of-sight photospheric magnetic field measured prior to the
flare, at 04:51:30 UT, by SOHO/MDI in the optical band using
Zeeman line splitting, where grades of yellow (blue) indicate
positive (negative) values. Spatial distributions of the total and
polarized intensities at 17 GHz are shown using various contours
as indicated in the plot at the levels of 40, 70, and 95\% of the
corresponding peak value. Right panel shows spatial distribution
of the positively polarized 17 GHz intensity (same as solid red
contours in the left panel) in grades of red with the 34 GHz
 total intensity superimposed in green contours at 40, 70, and
 95\% of the peak value. This spatial relationship, together with
 Fig. \ref{fig:Fig2}, prove that the emissions at 17 and 34 GHz,
 coming from the same spatial location, have opposite circular
 polarization as predicted for the electron/positron combination
 as in Fig. \ref{fig:Fig1}.}\label{fig:Fig3}
 \end{figure}

Figure \ref{fig:Fig2} displays the microwave light curves of the
event in intensity and circular polarization. Remarkably, the
polarization at 35 GHz is opposite to that at 17 GHz as expected
for the positron contribution. Then, to check this hypothesis we
have to examine the spatially resolved microwave and magnetic
field measurements. Figure \ref{fig:Fig3} clearly shows that (i)
there is only a single source at 17 and 34 GHz; (ii) the microwave
source at 34 GHz spatially coincides with that at 17 GHz and also
with the area where the 17 GHz emission is uniformly RCP; and
(iii) the northern end of the 17 GHz RCP source projects onto
positive (North) magnetic polarity, while the 17 GHz LCP source
projects onto negative (South) magnetic polarity. These regions
indicate photospheric footpoints presumably connected by a coronal
magnetic flux tube \citep{Bastian} illuminated by the microwave
brightness. Then, the region occupied by the 17 GHz RCP emission
(red solid contours in the left panel or grades of red in the
right image) traces the fraction of this loop with the positive
line-of-sight magnetic field. These, along with Fig.
\ref{fig:Fig2}, confirm that radiation at 17 GHz is polarized as
the extraordinary wave mode (as expected for electrons) and so
emission at 34 GHz is polarized as the ordinary wave mode (as
expected for positrons); thus, the conclusion that the observed
microwave emission at 34 GHz is produced by fast flare-created
positrons is inescapable for the event analyzed as there is no simple way of interpreting the reported observation with only an electron component.
As has been said
this event is not unique. We found similar evidence for
relativistic positrons in five other solar flares.

\section{Discussion}

This finding is fundamentally important for studying astrophysical
energy release and nuclear interactions. Indeed, understanding of
charged particle acceleration in astrophysical sources with strong
energy release -- including gamma--ray bursts, supernova remnants,
and solar flares -- requires complementary knowledge about both
fast electrons and ions (\cite{Lin03}, \cite{Shih}, \cite{Lin11},
\cite{Forrest}). At present, however, the information about the
accelerated ions is highly limited and comes primarily from the
gamma-ray lines produced at nuclear de-excitations \citep{Hurford}
or high-energy gamma-ray emission produced from $\pi^0$ decay and
associated gamma-ray continuum produced by secondary positrons
(e.g., \cite{Ackermann12}, \yearcite{Ackermann2013a},
\yearcite{Ackermann2013b}).

Another, potentially highly informative, but almost unexplored
channel is through the microwave emission produced by secondary
positrons created either from $\beta^+$ decaying radioactive
nuclei or from decay of $\pi^+$ particles. So far these positrons
were observed via the positron annihilation line at 0.511 MeV
(\cite{Murphy}, \cite{Share}, \cite{Schrijver}). Although helpful,
the diagnostics potential of this line is highly limited as it is
produced by already thermalized particles; thus, no information
about the original positron energy spectrum or transport can be
obtained from the annihilation line. Diagnostics based on the
high-energy bremsstrahlung produced by the secondary positrons is
potentially more promising. However, the ambiguity of the
leptonic-vs-hadronic spectral fits to the gamma-ray spectra (e.g.,
\cite{Ackermann12}) introduces an uncertainty to this gamma-ray
diagnostics.

In contrast, the discovery of the microwave emission produced by
flare-created relativistic positrons opens a new window for
studying the nuclear component of the accelerated particles and
their interactions. Indeed, until recently only the largest,
X-class flares were considered as potentially efficient nuclear
accelerators. However, Fermi LAT observations (e.g.,
\cite{Ackermann2013a}) detected the high-energy $\pi^0$-created
gamma-ray component from much weaker, M- and C- class flares.
Thus, detection of the nuclear acceleration signature depends on
the available sensitivity of the observational tool, rather than
on the X-ray class of the flare. The microwave observations with
their extraordinary sensitivity to small numbers of emitting
particles promise  a new  look at relativistic positrons and,
accordingly,  the acceleration and transport of nuclei in solar
flares.

Coming years
will bring an opportunity to routinely detect the relativistic
positron contributions in flares using new radio facilities,
including Jansky VLA \citep{Perley} and ALMA (\cite{Beasley},
\cite{Karlický}), having an ability to image circular polarization
at the relevant high microwave/mm frequencies. This will provide
us with invaluable information on the relativistic positron
spectra, spatial distribution, and evolution in solar flares, and
help to much better constrain the nuclear component of the
flare-accelerated particles.


\bigskip

Acknowledgements. We thank Dale Gary and Victoria Kurt for valuable comments. This
work was supported in part by NSF grants AGS-0961867, AST-0908344,
AGS-1250374 and NASA grants  NNX11AB49G and NNX13AE41G to New
Jersey Institute of Technology. This study was supported by the
Russian Foundation of Basic Research (12-02-91161, 12-02-00173,
12-02-10006) and by a Marie Curie International Research Staff
Exchange Scheme Fellowship within the 7th European Community
Framework Programme. The work is supported in part by the grants
of Ministry of education and science of the Russian Federation
(State Contracts 16.518.11.7065 and 02.740.11.0576). We thank the
Nobeyama team and SOHO team for providing open access to their
data. This work also benefited from workshop support from the
International Space Science Institute (ISSI). 


\end{document}